\documentstyle{article}
\begin{document}
\newcommand{\beq}{\begin{equation}}
\newcommand{\eeq}{\end{equation}}
\newcommand{\beqn}{\begin{eqnarray}}
\newcommand{\eeqn}{\end{eqnarray}}
\newcommand{\dpf}{\displaystyle\frac}
\newcommand{\no}{\nonumber}
\newcommand{\ep}{\epsilon}
\begin{center}
{\Large Two kinds of extreme black holes and their classification} 
\end{center}
\vspace{1ex}
\centerline{\large Bin Wang\footnote[0]{e-mail:binwang@fudan.ac.cn},
\ Ru-Keng Su\footnote[0]{e-mail:rksu@fudan.ac.cn}}
\begin{center}
{China Center of Advanced Science and Technology (World Laboratory),
P.O.Box 8730, Beijing 100080, P.R.China \\
and \\
Department of Physics, Fudan University, Shanghai 200433, P.R.China}
\end{center}
\vspace{6ex}
\begin{abstract}
%\vspace{1ex}
%\begin{minipage}{130mm)
%\end{minipage}
According to different topological configurations, we suggest that there
are 
two kinds of extreme black holes in the nature. We find that the Euler
characteristic
plays an essential role to classify these two kinds of extreme black holes.
For the 
first kind of extreme black holes, Euler characteristic is zero, and for
the second kind, 
Euler characteristic is two or one provided they are four dimensional holes
or two dimensional holes respectively.
\end{abstract}
\vspace{6ex}
\hspace*{0mm} PACS number(s): 04.70.Dy, 04.20.Gz, 04.62.+v.
\vfill
\newpage
Based upon the topological arguments between the
Reissner-Nordstr$\ddot{o}$m
(RN) extreme black hole (EBH) and nonextreme black hole (NEBH), Hawking et.
al.[1,2]
claimed that the EBH is a different object from its nonextreme counterpart
and the 
Bekenstein-Hawking (BH) formula of the entropy fails to describe the
entropy 
of EBH. A RN EBH has zero entropy, despite the nonzero area of the event
horizon. Their 
result had been extended to two-dimensional black holes [3].

Contrary to the above results, starting from a grand canonical ensemble,
Zaslavskii
shew that a RN black hole can approach the extreme state as closely as one
likes
in the topological sector of nonextreme configuration [4,5]. The
thermodynamical 
equilibrium can be fulfilled at every stage of this limiting process and
the BH 
formula of entropy is still valid for the final RN EBH. Zaslavskii also
found that the 
limiting geometry of RN black hole can be described by the
Bertotti-Robinson
(BR) spacetime.

According to their statements, the RN EBHs of Hawking et. al. and that of
Zaslavskii have very different properties. The first kind suggested by
Hawking
et. al. is the original EBH. It has zero entropy, infinite proper distance
$l$ between the horizon and any fixed point, in particular, purely extreme
topology. 
This kind of RN EBH cannot be formed in gravitational collapse, or
assimilating 
infalling charged particle and shell [6] from nonextreme RN black hole. It
can 
only arise through pair creation from the beginning of the universe [1].
The second 
kind suggested by Zaslavskii is a quite different object which is got by
first 
adopting the boundary condition $r_+=r_B$, where $r_+$ is the event horizon
and $r_B$ 
is the boundary of the cavity, and then the extreme condition $Q=M$. It's
entropy
satisfies BH formula like NEBH. It has finite proper distance $l$ and is
still 
in the topological sector of nonextreme configuration. These results
naturally
lead one to an impression that there are two kinds of RN EBHs, both
satisfying 
the extreme condition but with different characters in the nature. It is of
interest 
to ask whether these results can be extended to more general, not only
include 
RN black hole, but also include other black holes (at least include
two-dimensional 
(2D) or four-dimensional (4D) black holes). Are there two kinds of EBH in
the nature?
If there are, how to classify them? This paper envolves from an attempt to
answer
this problem and give a classification to two kinds of general extreme
holes.

Since the close relation between the black hole intrinsic thermodynamics
and its topology[7,8]
and the topological difference between the first kind and second kind EBHs,
it is natural 
to address this problem from the beginning by their topological characters.
We will prove the topological properties play an essential role in the
classification
of these two kinds of EBHs.

We study the four-dimensional (4D) black holes first.

I. 4D black holes.

The Euclidean metric form of 4D spherical black holes reads
\beq
{\rm d}s^2=e^{2U(r)}{\rm d}t^2+e^{-2U(r)}{\rm d}r^2+R^2{\rm d}\Omega^2
\eeq
Using the Gauss-Bonnet (GB) theorem and the boundary condition, one finds
the GB action
\beqn
S_{GB} & = & \dpf{1}{32\pi^2}\int_M\varepsilon_{abcd}R^{ab}\wedge R^{cd} \\
\no
       & = &\dpf{1}{4\pi^2}(\int_{\partial V}-\int_{\partial
M})\omega^{01}\wedge R^{23}
\eeqn
and the topological parameter, the Euler characteristic $\chi$ takes the
form[7,8]
\beq
\chi=\dpf{\beta}{2\pi}[(2U'e^{2U})(1-e^{2U}R'^2)]^{r_0}_{r_+}
\eeq
where $\beta=4\pi[(e^{2U})'_{r_+}]^{-1}$ is the inverse temperature.

A. RN black hole

The metric is in the form of Eq(1), but
\beq
e^{2U(r)}=1-\dpf{2M}{r}+\dpf{Q^2}{r^2},  R^2=r^2
\eeq
The extreme case corresponds to $M=Q$.  As was pointed out by Ref[1], for
the first 
kind of extreme RN hole, due to infinitely far away of the horizon
location, 
there are no conical singularities, which corresponds to no fixing
imaginary time 
period $\beta$, then the Euler characteristic $\chi=0$[2]. For the second
kind 
of extreme RN black hole, the Euler characteristic has not been calculated
before. 
Directly applying Eq(3) and adopting the boundary condition limit
($r_+\rightarrow r_B$)
first and then the extreme condition limit ($M\rightarrow Q$) afterwards,
we find
\beqn
\chi & = & [\dpf{\beta}{\pi
r^3}(M-\dpf{Q^2}{r})(2M-\dpf{Q^2}{r})]_{r=r_+=r_B}\vert_{extr} \\ \no
     & = & \dpf{4\pi r^6_B(r_B-M)}{2\pi r^6_B(r_B-M)}\vert_{extr}=2
     \eeqn
where we have considered $\beta=\dpf{4\pi r_+^3}{r_+^2-Q^2}$, and substract
the 
influence of the asymptotically flat spaces as in ref.[7,8]. The value for
the second kind of extreme RN hole is in agreement with that of the
nonextreme cases.

B. 4D dilaton black hole.

We extend our discussions to a more general case, namely, the dilaton black
hole.
The metric is still in the form of Eq(1), but
\beqn
e^{2U} & = & (1-\dpf{r_+}{r})(1-\dpf{r_-}{r})^{(1-a^2)/(1+a^2)} \\ \no
R^2 & = & r^2[1-\dpf{r_-}{r}]^{2a^2/(1+a^2)} \\ \no
2M & = & r_++\dpf{1-a^2}{1+a^2}r_- \\ \no
Q^2 & = & \dpf{r_+r_-}{1+a^2}
\eeqn
this solusion reduces to the RN case when $a=0$ and corresponds to the
black hole 
obtained from string theory [9] when $a=1$. We focus our attention on the
case 
$0<a\leq 1$.

The Euler characteristic $\chi$ for the first kind of EBH has been
calculated in
[7] and got $\chi=0$. Due to nonzero $[(e^{U}R')^2]_{extr}\vert_{r=r_+}$, 
if we first take extreme limit and then approach the horizon, some pecular
outcome will emerge.
To overcome this difficulty and obtain a unique and satisfactory result of
$\chi $, they add an inner boundary $r_0=r_++\ep$ and set $\ep\rightarrow
0$ at 
the end of calculation.

However, for the second kind of extreme dilaton hole, if taking boundary
limit
($r=r_+=r_B$) first and then imposing the extreme condition ($r_+=r_-$)
afterwards, 
we find $[(e^{U}R')^2\vert_{r=r_+=r_B}]_{extr}=0$, and 
\beqn
\chi & = & [\dpf{\beta}{2\pi}2U'e^{2U}]_{r_+=r_B}\vert_{extr} \\ \no
     & = & \dpf{4\pi r_B}{2\pi
r_B}[\dpf{r_B(r_B-r_-)}{r_B(r_B-r_-)}]^{(1-a^2)/(1+a^2)}\vert_{extr}=2
\eeqn
This value is in consistent with that of the nonextreme dilaton case[8].

C. Kerr black hole.

The metric of the Kerr black hole reads
\beqn
{\rm d}s^2 & = & -\dpf{\triangle}{\Sigma^2}[{\rm d}t-a\sin^2\theta{\rm
d}\phi]^2
+\dpf{\sin^2\theta}{\Sigma^2}[(r^2+a^2){\rm d}\phi-a{\rm d}t]^2 \\ \no
&   & +\dpf{\Sigma^2}{\triangle}{\rm d}r^2+\Sigma^2{\rm d}\theta^2
\eeqn
where 
\beq
\triangle=r^2-2Mr+a^2, \Sigma^2=r^2+a^2\cos^2\theta, a=J/M
\eeq
$M,J$ are the mass and angular momentum. The event horizon and the Cauchy
horizon locate at 
$r_+=M+\sqrt{M^2-a^2}, r_-=M-\sqrt{M^2-a^2}$, respectively. The extreme
case 
corresponds to $M=a$.

For the first kind of Kerr EBH, expending the metric coefficients near
$r=r_+$,
introducing $r-r_+=r_B\rho^{-1}$, we have
\beq
{\rm d}s^2=\Sigma_B^2\rho^{-2}\{-\dpf{r_B^2}{\Sigma_B^4}[{\rm
d}t-a\sin^2\theta{\rm d}\phi]^2
+\dpf{\rho^2\sin^2\theta}{\Sigma^4_B}[(r_B^2+a^2){\rm d}\phi-a{\rm
d}t]^2+{\rm d}\rho^2+\rho^2{\rm d}\theta^2\}
\eeq
The horizon satisfies the condition
\beq
\triangle=(r_B^2+a^2)f=r_B^2\rho^{-2}=0
\eeq
where
\beq
f=\dpf{(r-r_+)^2}{r^2+a^2}
\eeq
The horizon of the first kind of Kerr EBH locates at $\rho=\infty$. The
proper 
distance between $\rho=\infty$ and other $\rho<\infty$ is infinite. The
infinite 
horizon removes the conical singularity and makes the imaginary time
$\beta$ arbitrary.
Applying the arguments in [7,2], it leads unambiguously to $\chi=0$.

We now turn to discuss the second kind Kerr EBH. To get the Euler
characteristic,
we investigate the metric corresponding to this kind of Kerr EBH first.
Putting 
the nonextreme Kerr black hole in a cavity, the equilibrium condition is
\beq
\beta=\beta_0[f(r_B)]^{1/2},
T_0=1/\beta_0=\dpf{1}{4\pi}f'(r_+)=\dpf{\sqrt{M^2-a^2}}{2\pi(r_+^2+a^2)}
\eeq
where $f=\dpf{\triangle}{r^2+a^2}=\dpf{(r-r_+)(r-r_-)}{r^2+a^2}$
and $r_B$ is the cavity boundary. Choosing
\beq
r-r_+=4\pi T_0b^{-1}(\sinh^2x/2), b=\dpf{f''(r_+)}{2}
\eeq
and expending $f(r)=4\pi T_0(r-r_+)+b(r-r_+)^2$ near $r_+$, we find that in
the extreme limit $r_+=r_-=r_B (M=a), b=\dpf{1}{r_B^2+a^2}$, the metric 
becomes
\beqn
{\rm d}s^2 & = & -\Sigma_B^2\sinh^2x{\rm d}t^2_1+\Sigma^2_B{\rm
d}x^2+\Sigma^2_B{\rm d}\theta^2 \\ \no
           &   & +\dpf{\sin^2\theta}{\Sigma^2_B}[(r_B^2+a^2){\rm
d}\phi-a\sinh x\sqrt{r_B^2+a^2}{\rm d}t_1]^2
\eeqn
where time is normalized according to $t_1=2\pi T_0t, {\rm d}x=b{\rm
d}l^2$,
and $\Sigma^2_B=r^2_B+a^2\cos^2\theta$. Eq(15) is the generalization of the
Bertotti-Robinson 
(BR) spacetime [10] in the 4D nonspherical case.

The horizon of the extreme Kerr black hole can be detected by 
\beq
\triangle=(r_B^2+a^2)f=(r_B^2+a^2)\dpf{f'(r_+)}{4}(b^{-1}\sinh^2x)=0
\eeq
which locates at finite $x$, (say $x=0$). So the proper distance between
the 
horizon and any other fixed point is finite. By means of the formula of
$\chi$[8] and 
the extreme condition $(r_+=M=a)$, we obtain
\beqn
\chi & = & \dpf{Mr_+(r_+-M)}{4\pi^2}\int^{\beta_0}_0{\rm
d}\tau\int^{2\pi}_0{\rm d}\phi\int^{\pi}_0
\dpf{(r_+^2-3M^4\cos^4\theta)}{(r_+^2+M^2\cos^2\theta)^3}\sin\theta{\rm
d}\theta \\ \no
& = & \dpf{2}{\pi}\beta_0(r_+-M)\dpf{Mr_+}{(r_+^2+M^2)^2}=2
\eeqn
which is in agreement with that of the nonextreme case [8].

II. 2D black holes.

The formula of the Euler characteristic in 2D cases is [11]
\beq
\chi=\dpf{1}{2\pi}\int R_{1212}e^1\wedge e^2
\eeq

A. 2D charged dilaton black hole.

The metric of this black hole is [12,13]
\beqn
{\rm d}s^2 & = & -g(r){\rm d}t^2+g(r)^{-1}{\rm d}r^2 \\
g(r) & = & 1-2me^{-\lambda r}+ q^2e^{-2\lambda r} \\
e^{-2\phi} & = & e^{-2\phi_0}e^{\lambda r}, A_0=\sqrt{2}qe^{-\lambda r}
\eeqn
where $m$ and $q$ are the mass and electric charge of the black hole
respectively.
$m=q$ corresponds to the extreme case. Applying Eq(18) and substract the 
asymptotically flat space's influence, the Euler characteristic for the
NEBH reads
\beq
\chi=-\dpf{\beta}{2\pi}[-\lambda me^{-\lambda r}+\lambda q^2e^{-2\lambda
r}]_{r_+}=1
\eeq
where $1/{\beta}=g'(r_+)/4\pi$ [12].

We now extend above calculation of $\chi$ to two kinds of EBHs. For the
first 
kind of EBH, taking the extreme condition first
\beq
\chi=-\dpf{1}{2\pi}\beta_0[-m\lambda e^{-\lambda r}+m^2\lambda e^{-2\lambda
r}]_{r_+}
\eeq
and considering for the original EBH $r_+=\dpf{1}{\lambda}\ln m$, we have 
\beq
\chi=0
\eeq
But for the second kind of EBH, adopting the boundary condition first and
then 
the extreme condition, we obtain
\beqn
\chi & = & -\dpf{1}{2\pi}\beta_0[-m\lambda e^{-\lambda r}+q^2\lambda
e^{-2\lambda r}]_{r_+=r_B}\vert_{extr} \\ \no
& = &
\dpf{2\pi(m+\sqrt{m^2-q^2})\lambda\sqrt{m^2-q^2}}{2\pi\lambda\sqrt{m^2-q^2}(
m+\sqrt{m^2-q^2})}\vert_{extr}=1
\eeqn
We find the same result as that of the NEBH.

B. 2D Lowe-Strominger black hole.

The metric has the same form as Eq(19), but [14]
\beqn
g(r) & = & \lambda^2r^2-m-\dpf{J^2}{4r^2} \\
A_0 & = & -\dpf{J}{2r^2} \\
e^{-2\phi} & = & r
\eeqn
Using Eq(18), the Euler characteristic for this NEBH is 
\beq
\chi=-\dpf{\beta_0}{2\pi}[-\lambda^2 r+\dpf{J^2}{4r^3}]_{r_+}=1
\eeq
For the first kind of EBH, by taking the extreme limit $\lambda J=m$ first,

we find 
\beq
\chi=-\dpf{\beta_0}{2\pi\lambda^2}[-\lambda^4 r+\dpf{m^2}{4r^3}]_{r_+}
\eeq
and then using $r_+^2=\dpf{m}{2\lambda^2}$ for the original EBH, we get
\beq
\chi=0
\eeq
However, for the second kind of EBH, using the same treatment as before, we
find
\beq
\chi=-\dpf{\beta_0}{2\pi}[-\lambda^2
r+\dpf{J^2}{4r^3}]_{r_+=r_B}\vert_{extr}=1
\eeq
which agrees to Eq(29) of the NEBH.

According to above calculations, in summary, we suggest that there are two
kinds of 
4D and 2D EBHs in the nature. The first kind of EBHs is the original EBH.
They can be
obtained in mathematics by first taking the extreme limit and then the
boundary 
limit. The entropy of this kind of EBH is zero. The second kind of EBHs
still 
holds the topological configuration of NEBH. They can be obtained in
mathematics by
doing the other way round, i.e. taking the boundary limit first and the
extreme limit 
afterwards. The entropy of this kind of EBHs satisfies the BH formula.
These 
two kinds of EBHs have different intrinsic thermodynamics. We have shown
that the
Euler characteristic plays an essential role to classify these two kinds of
EBHs. For the first kind, 
Euler characteristic is zero; and for the second kind, the Euler
characteristic 
equals to two or one provided they are 4D or 2D EBHs respectively.

\hspace{0mm} This work was supported in part by NNSF of China. B.Wang would
like to thank helpful discussions with Prof. Randjbar when he visited ICTP.


\vfill
\newpage
\begin{thebibliography}{x}
\bibitem{} S.W.Hawking, G.Horowitz and S.Ross,
           Phys. Rev. D 51, 4302 (1995)
\bibitem{} C.Teitelboim,
           Phys. Rev. D 51, 4315 (1995)
\bibitem{} A.Kumar and K.Ray, Phys. Rev. D 51, 5954, (1995)
\bibitem{} O.B.Zaslavskii, Phys. Rev. Lett. 76, 2211 (1996)
\bibitem{} O.B.Zaslavskii, Phys. Rev. D 56, 2188 (1997)
\bibitem{} B.Wang, R.K.Su, P.K.N.Yu and E.C.M.Young, Phys. Rev. D 57, 5284
(1998) 
\bibitem{} G.W.Gibbons and R.E.Kallosh, Phys. Rev. D 51, 2839 (1995)
\bibitem{} S.Liberati and G.Pollifrone, Phys. Rev. D 56, 6468 (1997)
\bibitem{} D.Garfinkle, G.T.Horowitz and A.Strominger, Phys. Rev. D 43,
3140 (1991)
\bibitem{} B.Bertotti, Phys. Rev. 116, 1331 (1959) \\
           I.Robinson, Bull. Acad. Pol. Sci. 7, 351 (1959)
\bibitem{} T.Eguchi, P.B.Gikey and A.J.Hanson, Phys. Rep. 66, 6 (1980)
\bibitem{} C.Nappi and A.Pasquinucci, Mod.Phys.Lett.A 7, 3337 (1992)
\bibitem{} M.D.McGuigan, C.R.Nappi and S.A.Yost, Nucl. Phys. B 375, 121
(1992)
\bibitem{} D.Lowe and A.Strominger, Phys. Rev. Lett 73, 1468 (1994)
\end{thebibliography}
\end{document}